\documentclass[10pt,journal,compsoc]{IEEEtran}
\usepackage{amsmath,amsfonts}
\usepackage{algorithmic}
\usepackage{makecell}
\usepackage{algorithm}
\usepackage{array}
\usepackage{booktabs}
\usepackage{tabularx}
\newcolumntype{R}{>{\raggedleft\arraybackslash}X}
\usepackage{multirow}
\usepackage{subcaption}
\usepackage{textcomp}
\usepackage{stfloats}
\usepackage{url}
\usepackage{verbatim}
\usepackage{graphicx}
\usepackage[nocompress]{cite}
\usepackage{enumitem}

\begin{document}

\bstctlcite{IEEEexample:BSTcontrol}

\title{ADD 2023: Towards Audio Deepfake Detection and Analysis in the Wild}

\author{Jiangyan~Yi,~\IEEEmembership{Member,~IEEE,}
  Chu~Yuan~Zhang,
  Jianhua~Tao,~\IEEEmembership{Senior~Member,~IEEE,}
  Chenglong~Wang,
  Xinrui~Yan,
  Yong~Ren,
  Hao~Gu,
  Junzuo~Zhou
\IEEEcompsocitemizethanks{\IEEEcompsocthanksitem Jiangyan~Yi and Chu~Yuan~Zhang contributed equally.
\IEEEcompsocthanksitem Jiangyan~Yi,
Chu~Yuan~Zhang,
Chenglong~Wang,
Xinrui~Yan,
Yong~Ren,
Hao~Gu,
and Junzuo~Zhou are with the Institute of Automation, Chinese Academy of Sciences, Beijing 100190, China. \protect\\
Email: jiangyan.yi@nlpr.ia.ac.cn
\IEEEcompsocthanksitem Jianhua Tao is with the Department of Automation, Tsinghua University, Beijing 100190, China.}%
}

\markboth{Journal of \LaTeX\ Class Files,~Vol.~14, No.~8, August~2021}%
{Yi \MakeLowercase{\textit{et al.}}: ADD 2023: Towards Audio Deepfake Detection and Analysis in the Wild}


\IEEEtitleabstractindextext{%
\begin{abstract}
  The growing prominence of the field of audio deepfake detection is driven by its wide range of applications, notably in protecting the public from potential fraud and other malicious activities, prompting the need for greater attention and research in this area. The ADD 2023 challenge goes beyond binary real/fake classification by emulating real-world scenarios, such as the identification of manipulated intervals in partially fake audio and determining the source responsible for generating any fake audio, both with real-life implications, notably in audio forensics, law enforcement, and construction of reliable and trustworthy evidence. To further foster research in this area, in this article, we describe the dataset that was used in the fake game, manipulation region location and deepfake algorithm recognition tracks of the challenge. We also focus on the analysis of the technical methodologies by the top-performing participants in each task and note the commonalities and differences in their approaches. Finally, we discuss the current technical limitations as identified through the technical analysis, and provide a roadmap for future research directions. The dataset is available for download at \url{http://addchallenge.cn/downloadADD2023}
\end{abstract}

}

\maketitle
\IEEEdisplaynontitleabstractindextext
\IEEEpeerreviewmaketitle

\IEEEraisesectionheading{\section{Introduction}\label{sec:introduction}}
\IEEEPARstart{R}{ecent} rapid advancements in text-to-speech (TTS)~\cite{tan2021survey} and voice conversion (VC)~\cite{sisman2020or} technologies over the past decades have made it possible to generate high-quality and realistic audio that can be difficult to distinguish from real audio with the naked ear. Such a technology has a potential to be abused and misused, notably in generating deepfake audio for impersonation, fraud, and other malicious purposes. The rapid progress in TTS and VC technologies means that these attacks can be readily launched by anyone with a computer and a microphone, and rapidly spread through social media. Therefore, it is urgent to develop effective deepfake audio detection methods to protect the public from being deceived by deepfake audio.

In response to this necessity, the Automatic Speaker Verification Spoofing and Countermeasures (ASVspoof) challenge~\cite{2021ASVspoof} in 2021 and 2024, notably included a speech deepfake detection (DF) track which specifically focused on the detection of VC and TTS from audio, which further spurs research in this area. The first Audio Deepfake Detection Challenge (ADD 2022)~\cite{Yi2022ADD} was organized 
to further promote research on audio deepfake detection. 
Yet, current efforts on deepfake audio detection focus on the binary classification of real and fake audio, which may sometimes prove insufficient in real-world scenarios. For instance, in the case of audio forensics and law enforcement, it is often crucial to identify the specific intervals within partially fake audio where manipulation occurs; determining the source of the fake audio is also important for attribution and accountability, protecting intellectual property rights, and preventing the spread of misinformation. This improved detection and analysis of deepfake audio is also essential for constructing reliable and trustworthy evidence, not only in court, but also in other areas such as journalism and social media. Hence, greater attention is needed to advance deepfake audio detection beyond binary real/fake classification. These challenges represent the next frontier in combating audio manipulation and deception.

In response to this growing need to advance deepfake audio detection research in this direction, we launched a second Audio Deepfake Detection Challenge (ADD 2023)\footnote{\url{http://addchallenge.cn/add2023}} to further promote research on deepfake audio detection and analysis. The tasks in the ADD 2023 challenge as well as the data used in each task are designed to emulate real-world scenarios and to motivate research that goes beyond the traditional binary classification of real and fake audio and to further accelerate and foster research on detecting and analysing deepfake audio. We hoped that the insight gained from our analysis will help further advance the research on deepfake audio detection and analysis, and that the ADD 2023 challenge will serve as a stepping stone for future research in this area.

With the successful conclusion of the challenge, in order to further prompt relevant research in audio deepfake detection, we release a dataset for the ADD 2023 challenge, partitioned into four subsets, each corresponding to a task in the challenge.  We break down the dataset into four tasks and provide a detailed description of each dataset in Section~\ref{sec:datasets}. This dataset is designed to simulate different acoustic environments, adversarial attacks, and manipulation techniques, allowing participants' models to be tested under conditions that resemble those found in real-life scenarios, whether due to the rapid advancement in audio deepfake technologies, social media uploads, or unknown deepfake algorithms.

Furthermore, we provide a thorough analysis of the top-performing systems in each track in order to identify key strategies and common techniques employed by successful teams. This analysis, presented in Section~\ref{sec:challenge-results}, offers valuable insights into the current popular methodologies in the field of deepfake audio detection and analysis, and lays the groundwork for future research. Finally, we discuss the limitations observed in the current approaches and outline future directions for advancing research in deepfake audio detection and analysis in Section~\ref{sec:limitations}, highlighting the ongoing need for innovation and collaboration in this rapidly evolving area.

The primary contributions of this article include: 
\begin{enumerate}[leftmargin=*,nosep,label=(\arabic*)]
  \item We detail the datasets used in all tracks of the ADD 2023 challenge and provide a comprehensive description of each of the datasets, which have been made publicly available for download;
  \item We analyze the technical methodologies used by the top-performing participants in each task of the ADD 2023 challenge and note the commonalities and differences in their approaches;
  \item We discuss the current technical limitations identified during the challenge and provide a roadmap for future research directions.
\end{enumerate}


\begin{figure*}
  \centering
  \begin{subfigure}{0.65\linewidth}
    \centering
    \includegraphics[width=\linewidth]{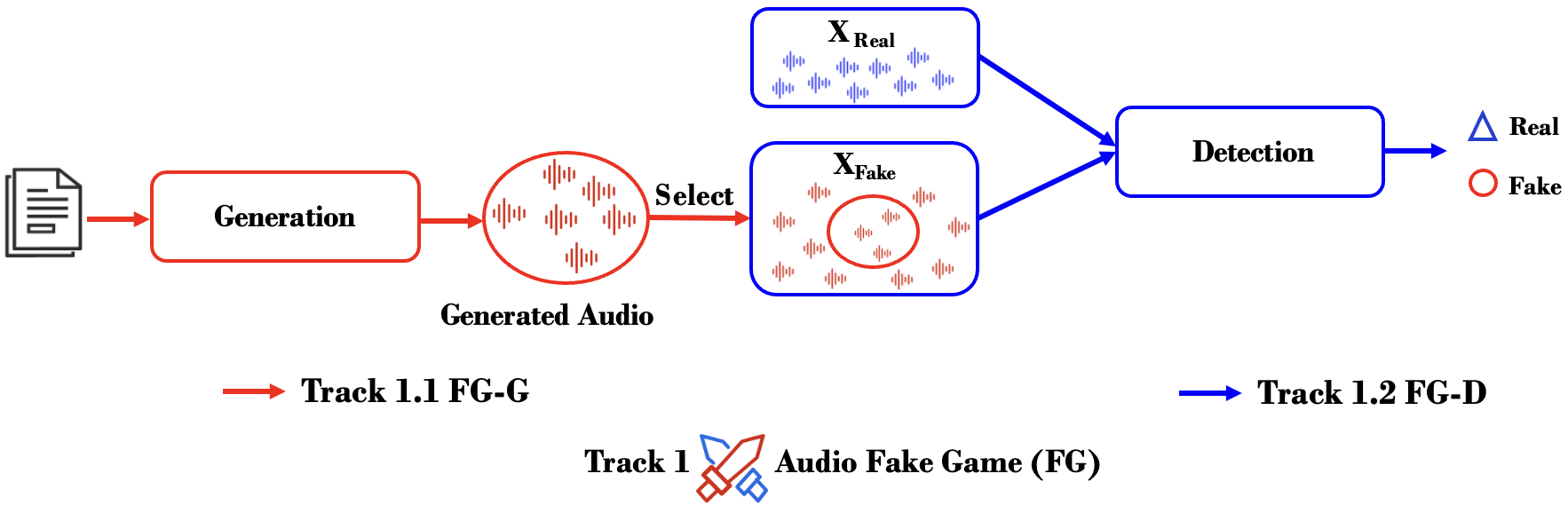}
    \label{fig:dataset-overall}
  \end{subfigure}

  \vspace{-2\baselineskip}

  \begin{subfigure}[b]{0.63\linewidth}
    \centering
    \includegraphics[width=\linewidth]{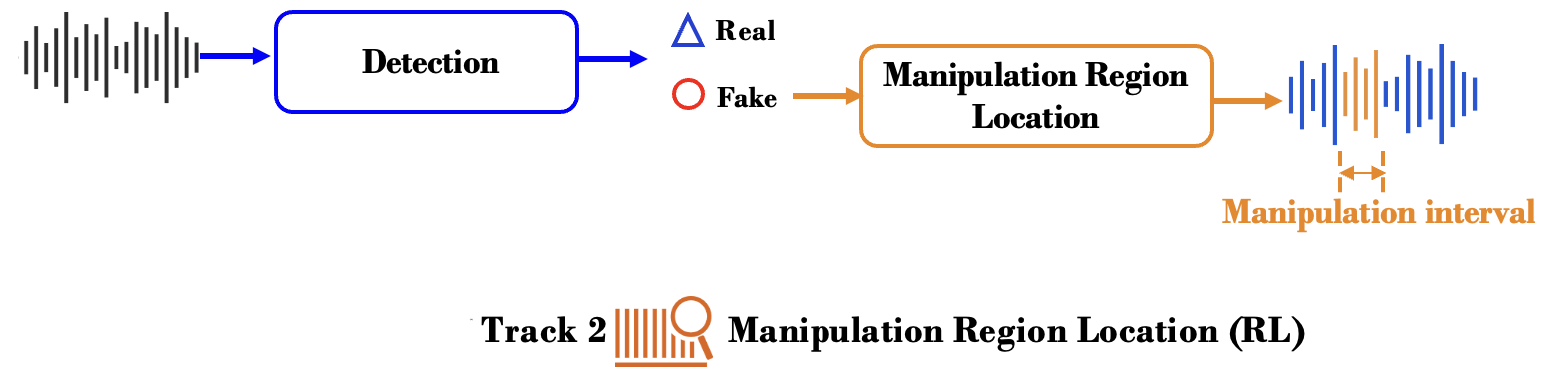}
    \label{fig:track-2:fake-region-length}
  \end{subfigure}
  \hfill
  \begin{subfigure}[b]{0.36\linewidth}
    \centering
    \includegraphics[width=\linewidth]{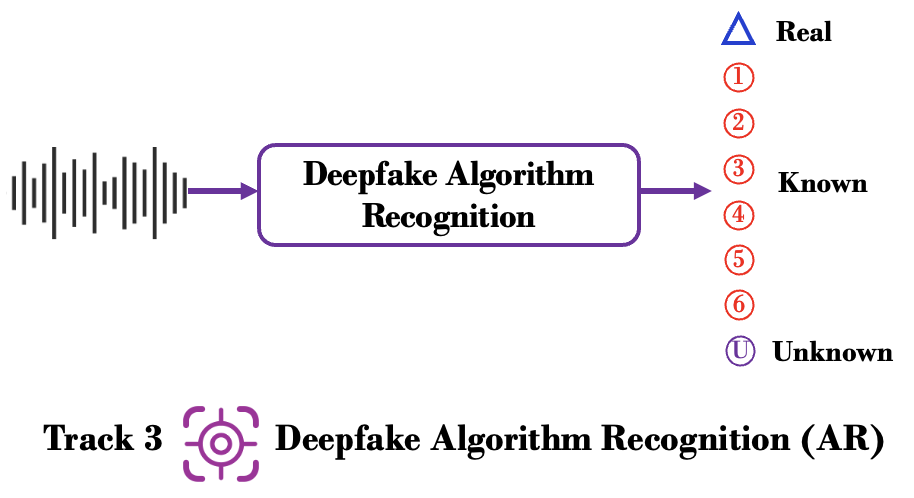}
    \label{fig:track-2:fake-region-length}
  \end{subfigure}
  \caption{The three tracks of ADD 2023.}\label{fig:dataset-overall}
\end{figure*}

\section{Challenge Outline}\label{sec:challenge-outline}

In this section, we outline the ADD 2023 challenge, which was divided into three tracks (see Figure~\ref{fig:dataset-overall}):
\begin{itemize}
  \item \textbf{Track 1: Audio fake game (FG)}, representing the attack-and-defense game~\cite{Peng2021DFGC} between the attacking party, which was tasked with generating deepfake audio, and the defending party, which was tasked with detecting deepfake audio. This track was further divided into two different yet interconnected sub-tracks, each with two rounds of evaluations, allowing for adaptation and evolution of the participants' systems:
  \begin{itemize}
    \item \textbf{Track 1.1: Generation task (FG-G)}, representing the attacking party in the process. The participants were tasked with generating deepfake audio that convincingly emulated the characteristics of genuine audio.
    \item \textbf{Track 1.2: Detection task (FG-D)}, representing the defending party in the process. The participants were tasked with detecting deepfake audio, some of which were generated by the participants in the FG-G sub-track, and distinguishing it from genuine audio.
  \end{itemize}
  \item \textbf{Track 2: Manipulation region location (RL)}, representing the task of locating the specific intervals within partially fake audio where manipulation occurs. The manipulation region, for the purpose of the challenge, is defined as the interval of the audio signal that is replaced by a different audio signal that is either generated with the same target speaker ID, or from a different genuine recording of the same speaker.
  \item \textbf{Track 3: Deepfake algorithm recognition (AR)}, representing the task of determining the source algorithm responsible for generating a given piece of fake audio, as well as an unknown deepfake algorithm.
\end{itemize}

\section{Evaluation Metrics}\label{sec:metrics}

This section presents the evaluation metrics for the four tasks of the ADD 2023, measuring participants' performance various tasks. The precise metrics used for each task are reported in \cite{yi2023add2023}.

\subsection{Track 1.1: Generation task (FG-G)}\label{ssec:metrics:track-1-1}

The evaluation metric for the generation task is the deception success rate (DSR), based on the percentage of utterances misclassified as genuine by the FG-D sub-track detection systems. In the second round, participants must also deceive a target detection system, with performance scored by the weighted sum of detection errors from both the target model and FG-D sub-track models. Each team's score is the weighted sum of DSRs from both rounds, emphasizing the second round to encourage adapting methods.

\subsection{Track 1.2: Detection task (FG-D)}\label{ssec:metrics:track-1-2}

Given the task of Track 1.2 in distinguishing genuine and generated audio samples, the evaluation metric used for this sub-track is the equal error rate (EER). The overall performance of detection systems is evaluated as the weighted average EER (WEER) of the detection systems from both rounds, allowing for more emphasis on the second round.

\subsection{Track 2: Manipulation region location (RL)}\label{ssec:metrics:track-2}

The evaluation metric for the RL track is the sentence-level accuracy ($A_s$) and the segment-level $\text{F}_1$ score ($F_{1s}$). The overall performance of each team in the RL track (Score)
is thus evaluated as a weighted sum of $A_s$ and the $F_{1s}$, with more emphasis on the latter, which is more challenging to achieve and also the main objective of this track, without ignoring the overall $A_s$, which is also important in real-world applications.

\subsection{Track 3: Deepfake algorithm recognition (AR)}\label{ssec:metrics:track-3}

The performances of the participants' systems in Track 3 are evaluated using the macro-averaged $\text{F}_1$ score~\cite{yan2022initial} ($F_1$), which includes both known and unknown class samples, testing the generalization capabilities of the participants' systems.

\section{Datasets}\label{sec:datasets}

In this section, we present and describe the datasets\footnote{\url{http://addchallenge.cn/downloadADD2023}} used in the ADD 2023 challenge. Each dataset is used for a specific task that corresponds to a track in the challenge.

In order to better represent real-world scenarios, the datasets are designed according to the following general principles: first, there should be a noticeable difference in the training and testing set data distributions, to ensure that models are able to generalize well to unseen data present in real-life applications; second, there needs to be a sense of attack-and-defence dynamic, which is especially important for the FG track; and third, the inclusion of unknown deepfake algorithms and manipulation techniques aims to further test the robustness of the participants' systems.

\subsection{Track 1.1: Generation task (FG-G)}\label{ssec:challenge-outline:fg-g}

The FG-G track being focused on the generation of deepfake audio, its dataset is composed of speech corpora with transcripts. In this case, we use the AISHELL-3 dataset~\cite{shi2020aishell3} in the training process, which is a large-scale multi-speaker Mandarin speech dataset for TTS that contains roughly 85 hours of speech data from 218 native Mandarin speakers, totalling 88,035 utterances. This is done to ensure that the generated audio is of high quality and is able to emulate the characteristics of genuine audio.


In addition, the two-round setup of the FG-G sub-track and its dataset is designed to encourage participants to adapt their methods and models to new data and conditions, as well as to test the robustness of their systems. In this sense, the release of a benchmark model for the second round of the FG-G sub-track is intended to provide a target for participants to deceive, and to further motivate the development of more advanced and sophisticated deepfake audio generation methods.

The testing set, summarized in Table~\ref{tab:datasets:track-1-1}, is composed of 998 text sentences (499 per each round) in Simplified Chinese characters, as well as four speaker IDs (two per each round, no overlap) from the aforementioned AISHELL-3 dataset.
The average length of these sentences (excluding punctuations) is 66.98 characters, with a maximum length of 147 characters, a minimum length of 39 characters, and a variance of 136.78 characters.

\begin{table}
  \centering
  \caption{Testing set for the FG-G sub-track (Track 1.1).}\label{tab:datasets:track-1-1}
  \begin{tabularx}{\linewidth}{Xrrrr}
    \toprule
    \multirow{2.5}{*}{\textbf{Round}} & \multicolumn{2}{c}{\textbf{\# Speakers}} & \multirow{2.5}{*}{\textbf{\# Sentences}} & \multicolumn{1}{c}{\textbf{\# Utterances}}\\ \cmidrule{2-3}
    & \textbf{\# Male } & \textbf{\# Female } & & \multicolumn{1}{c}{\textbf{Generated}} \\ \midrule
    \textbf{Round 1} & 1 & 1 & 499 & 998 \\
    \textbf{Round 2} & 1 & 1 & 499 & 998 \\
    \bottomrule
  \end{tabularx}
\end{table}

\begin{table}
  \centering
  \caption{Dataset for the FG-D sub-track (Track 1.2).}\label{tab:datasets:track-1-2}
  \begin{subtable}{\linewidth}
    \centering
    \caption{Training and development sets; the same used for both rounds.}\label{tab:datasets:track-1-2:train}
    \begin{tabularx}{\linewidth}{lXrrrr}
      \toprule[.75pt]
      \multicolumn{2}{c}{\multirow{2.5}{*}{\textbf{Origin}}} & \multicolumn{2}{c}{\textbf{Train}} & \multicolumn{2}{c}{\textbf{Dev}} \\ \cmidrule(lr){3-4}\cmidrule(lr){5-6}
      & & \multicolumn{1}{c}{\textbf{\# Spk}} & \multicolumn{1}{c}{\textbf{\# Utt}} & \multicolumn{1}{c}{\textbf{\# Spk}} & \multicolumn{1}{c}{\textbf{\# Utt}}  \\ \midrule[.5pt]
      \textbf{Real} &  AISHELL-3 & 60 & 3,012 & 60 & 2,307 \\\midrule
      \multirow{7}{*}{\textbf{Fake}} & HiFiGAN & 60 & 4,012 & 60 & 4,387 \\
      & LPCNet & 60 & 4,012 & 60 & 4,387 \\
      & Multiband MelGAN & 60 & 4,012 & 60 & 4,387 \\
      & StyleMelGAN & 60 & 4,012 & 60 & 4,387 \\
      & Parallel WaveGAN & 60 & 4,012 & 60 & 4,082 \\
      & World & 60 & 4,012 & 60 & 4,387 \\\addlinespace
      & \textbf{Total} & 360 & 24,072 & 360 & 26,027 \\
      \bottomrule[.75pt]
    \end{tabularx}
  \end{subtable}
  
  \vspace{\baselineskip}

  \begin{subtable}{\linewidth}
    \centering
    \caption{Testing sets. Note that the utterances for the two rounds are selected randomly and independently, and may have overlaps between them (except FG-G, where the two rounds are disjunct).}\label{tab:datasets:track-1-2:test}
    \begin{tabularx}{\linewidth}{lXrrrr}
      \toprule[.75pt]
      \multicolumn{2}{c}{\multirow{2.5}{*}{\textbf{Origin}}} & \multicolumn{2}{c}{\textbf{Round 1}} & \multicolumn{2}{c}{\textbf{Round 2}} \\ \cmidrule(lr){3-4}\cmidrule(lr){5-6}
      & & \multicolumn{1}{c}{\textbf{\# Spk}} & \multicolumn{1}{c}{\textbf{\# Utt}} & \multicolumn{1}{c}{\textbf{\# Spk}} & \multicolumn{1}{c}{\textbf{\# Utt}}  \\ \midrule[.5pt]
      \multirow{9}{*}{\textbf{Real}} &  AISHELL-3  & 50 & 3,012 & 50 & 2,307 \\
       & AISHELL-1& 200 & 10,000 & -- & -- \\
       & THCHS-30 & 30 & 7,000 & 30 & 7,000 \\
       & HI-MIA & 100 & 10,000 & 100 & 10,000 \\
       & Mobvoi & 100 & 20,000 & 100 & 20,000 \\
       & AliMeeting & 50 & 20,000 & 50 & 20,000 \\
       & Phone recording & 100 & 10,000 & 100 & 10,000 \\
       & ESD & -- & -- & 10 & 17,500 \\\addlinespace
        & \textbf{Total} & 630 & 80,012 & 440 & 86,807 \\
      \midrule
      \multirow{23}{*}{\textbf{Fake}} & HiFiGAN & 218 & 600 & 218 & 600 \\
      & LPCNet & 218 & 3,000 & 218 & 3,000 \\
      & Multiband MelGAN & 218 & 600 & 218 & 600 \\
      & Parallel WaveGAN & 218 & 600 & 218 & 600 \\
      & StyleMelGAN & 218 & 600 & 218 & 600 \\
      & World & 218 & 400 & 218 & 400 \\
      & WaveRNN & 218 & 3,600 & 218 & 3,600 \\
      & VITS & 218 & 1,334 & 218 & 1,334 \\
      & Grad-TTS & 218 & 1,333 & 218 & 1,333 \\
      & Speech Edit & 218 & 1,333 & 218 & 1,333 \\
      & Alibaba & -- & 482 & -- & 482 \\
      & DeepSound & -- & 400 & -- & 400 \\
      & iFlytek & -- & 400 & -- & 400 \\
      & Mobvoi & -- & 397 & -- & 397 \\
      & Baidu & -- & 387 & -- & 387 \\
      & Aispeech & -- & 385 & -- & 385 \\
      & DataBaker & -- & 400 & -- & 400 \\
      & Sogou & -- & 400 & -- & 400 \\
      & Tencent & -- & 400 & -- & 400 \\
      & Sohu & -- & 349 & -- & 349 \\
      & Blizzard & -- & 4,000 & --& 4,000 \\
      & FG-G & 2 & 11,976 & 2 & 11,976 \\
      \addlinespace
      & \textbf{Total} &  & 33,376 &  & 33,376 \\
      \bottomrule[.75pt]
    \end{tabularx}
  \end{subtable}
\end{table}

\subsection{Track 1.2: Detection task (FG-D)}\label{ssec:challenge-outline:fg-d}



The dataset used in the FG-D sub-track contains both real and fake speech, as shown in Table~\ref{tab:datasets:track-1-2}, with a sampling frequency of 16~kHz. Note that, while the speech samples are randomly selected for the two rounds of testing, there is no speaker overlap between the two rounds.

\subsubsection{Training and development sets}

The training and development sets of the FG-D track are deliberately designed to provide a more restricted set of data, reflecting the controlled and limited conditions under which detection models are typically trained and validated. This restriction is intentional, as it mirrors the reality that, during the training and validation phases, models often have access to a more homogenous and well-curated dataset. This dataset might lack the full diversity and complexity of the data that the model will encounter in practical, real-world applications.

In the training and development sets, only the AISHELL-3 dataset is used for genuine speech, totalling 60 speakers and 3,012 utterances for the training set and 60 speakers and 2,307 utterances for the development set (see Table~\ref{tab:datasets:track-1-2}(a)). For fake audio, we use samples generated from the following methods:
\begin{itemize}[leftmargin=*]
  \item \textbf{HiFiGAN~\cite{kong2020hifi}:} A GAN-powered model that generates high-fidelity audio from mel-spectrograms. 

  \item \textbf{LPCNet~\cite{valin2019lpcnet}:} A WaveRNN-based vocoder combining linear prediction coding and recurrent neural networks to generate high-quality audio.

  \item \textbf{Multiband MelGAN~\cite{yang2020multiband}:} A vocoder model based on MelGAN with larger receptive fields and multi-band discriminators.
  
  \item \textbf{StyleMelGAN~\cite{mustafa2021stylemelgan}:} A vocoder model based on MelGAN with temporal adaptive normalization and multiple random-window discriminators. 

  \item \textbf{Parallel WaveGAN~\cite{yamamoto2020parallel}:} A GAN-based vocoder incorporating non-autoregressive WaveNet~\cite{oord2016wavenet} for parallel and efficient waveform generation. 

  \item \textbf{World~\cite{morise2016world}:} A traditional vocoder system using spectral analysis to decompose the audio waveform into three main components: fundamental frequency (F0), spectral envelope, and non-periodic component. 
\end{itemize}

\subsubsection{Testing set}

For the testing set of FG-D dataset, we include audio samples from a wider range of sources for both real and fake audio (see Table~\ref{tab:datasets:track-1-2}(b)), since it is designed to be more challenging and diverse than the training and development sets, reflecting the broader range of scenarios that detection models are likely to encounter in real-world deployments, where the data may be more varied and unpredictable due to different environments, recording conditions, and deepfake manipulation techniques. The testing set, therefore, serves as a more challenging benchmark for evaluating the robustness and adaptability of the detection models.

The genuine audio portion of the Track 1.2 testing set comes from the following sources:
\begin{itemize}[leftmargin=*]
  \item \textbf{AISHELL-3~\cite{shi2020aishell3}:} A comprehensive and high-fidelity multi-speaker Mandarin speech corpus. We selected a set of 50 distinct speakers from this dataset, resulting in a combined recording duration surpassing 3.5~hours.
  \item \textbf{AISHELL-1~\cite{bu2017aishell1}:} A Mandarin speech corpus containing recordings from three different recording devices: high-fidelity microphones, Android smartphones, and iOS smartphones. We selected a total of 200 different speakers from this dataset, resulting in a combined recording duration exceeding 12~hours.
  \item \textbf{THCHS-30~\cite{wang2015thchs30}:} This dataset contains Chinese speech recordings that were recorded in a quiet office setting. We selected voice recordings from 30 distinct speakers, resulting in a cumulative duration of over 8~hours.
  \item \textbf{HI-MIA~\cite{qin2020himia}:} Originally a dataset for speaker verification containing recordings of 340 people in rooms designed for the far-field scenario. We have chosen a total of 100 speakers, resulting in a combined recording duration exceeding 11.5~hours.
  \item \textbf{Mobvoi~\cite{hou2019mobvoi}:} The MobvoiHotwords corpus consists of a collection of wake-up words collected from a commercial smart speaker by Mobvoi. It includes keyword and non-keyword utterances. We selected 100 different speakers from among them, resulting in a total duration exceeding 24 hours.
  \item \textbf{AliMeeting~\cite{Yu2022M2MeT}:} The AliMeeting dataset is a multi-speaker Mandarin speech corpus collected from real meetings, encompassing far-field audio captured using an 8-channel microphone array, in addition to near-field audio captured using individual participants' headset microphones. We have chosen 50 speakers from among them, resulting in a total duration exceeding 23 hours.
  \item \textbf{ESD~\cite{zhou2022esd}:} Contains speech by 10 native Chinese speakers, with each speaker contributing 350 utterances, covering 5 emotion categories, namely \emph{Neutral}, \emph{Happy}, \emph{Angry}, \emph{Sad}, and \emph{Surprise}. The average utterance and word duration are 3.22~s and 0.28~s respectively. All data is recorded in a typical indoor environment with an SNR of above 20~dB.
  \item \textbf{Phone recording:} We collected utterances from volunteers in ordinary day-to-day environments. These recordings were made using cellphone microphones and include 100 speakers, with a total duration exceeding 12 hours.
\end{itemize}
The fake audio portion of the dataset include audio samples generated from the same sources as the training and development sets, as well as several additional sources, roughly divided into three categories: TTS and VC models, external tools, and challenge participants. The sources include:
\begin{itemize}[leftmargin=*]
  \item \textbf{WaveRNN~\cite{kalchbrenner2018efficient}:} A single-layer recurrent neural network that predicts 16-bit audio waveforms, powered by gated recurrent units.

  \item \textbf{VITS~\cite{kim2021vits}:} An end-to-end TTS model based on conditional variational autoencoder with adversarial training.

  \item \textbf{Grad-TTS~\cite{popov2021gradtts}:} A diffusion-based generative model that can generate high-fidelity speech audio. The spectrogram generated by Grad-TTS is converted to waveform using HiFiGAN.
  
  \item \textbf{Speech Edit~\cite{wang2022campnet}:} A speech editing model employing a context-aware mask prediction network that can edit speech in the time domain.

  \item \textbf{Aliyun\footnote{\url{https://ai.aliyun.com/nls/tts}}:} An industry-leading voice synthesis solution from Alibaba, powered by the newest deep-learning technologies.
  
  \item \textbf{DeepSound\footnote{\url{https://www.deepsound.cn}}:} A TTS platform developed by DeepSound, providing a general speech synthesis solution that transforms text into expressive speech.
  
  \item \textbf{iFlytek\footnote{\url{https://global.xfyun.cn/products/text-to-speech}}:} A TTS platform developed by iFlytek, powered by the latest deep learning technologies. 
  
  \item \textbf{Mobvoi\footnote{\url{https://ai.chumenwenwen.com/}}:} A TTS platform developed by Mobvoi and powered by MeetVoice, an end-to-end speech synthesis engine that supports emotion synthesis.
  
  \item \textbf{Baidu\footnote{\url{https://ai.baidu.com/tech/speech/tts}}:} A deep learning-based TTS system developed by Baidu, powered by the latest deep learning technologies.

  \item \textbf{AISpeech\footnote{\url{https://beta.duiopen.com/openSource/technology/tts}}:} A leading voice synthesis solution in the Chinese industry. able to emulate various types of voices through its TTS and VC platform.
  
  \item \textbf{DataBaker\footnote{\url{https://www.data-baker.com/specs/compose/online}}:} Uses state-of-the-art TTS and VC platform based on the Transformer~\cite{vaswani2017attention}.
  
  \item \textbf{Sogou\footnote{\url{https://ai.sogou.com/product/tts}}:} A TTS and VC platform developed by Sogou based on a sequence-to-sequence acoustic model.
  
  \item \textbf{Tencent\footnote{\url{hhttps://www.tencentcloud.com/products/tts}}:} A TTS and VC platform developed by Tencent providing high quality speech synthesis services.
  
  \item \textbf{Sohu\footnote{\url{https://ai.sohu.com/}}:} A TTS and VC platform developed by Sohu.

  \item \textbf{Blizzard:} The fake audio samples generated by the participants in the Blizzard Challenge 2020~\cite{zhou2020blizzard}, with around 250 samples from each team, resulting in 4,000 utterances in total.

  \item \textbf{FG-G:} The fake audio samples generated by the participants in the FG-G sub-track of the ADD 2023 challenge, with 998 samples per team, totalling 11,976 utterances. Note that the two rounds of the FG-G sub-track are disjunct, \emph{i.e.}, the samples generated in round 1 are not included in round 2.
\end{itemize}

By designing the dataset in this way, we seek to emphasize the importance of generalization, encouraging researchers to develop models that can perform well not just on the curated datasets they are trained on, but also on the unpredictable and diverse data they will face in actual use cases. This approach is crucial for advancing the field of deepfake audio detection, as it pushes the development of models that are not only accurate in controlled settings but also resilient and effective in the dynamic, real-world scenarios where they are most needed.

\begin{table}
  \centering
  \caption{Dataset for the RL track (Track 2)}\label{tab:datasets:track-2}
  \begin{subtable}{\linewidth}
    \centering
    \caption{Training and development sets.}\label{tab:datasets:track-2:train}
    \begin{tabularx}{\linewidth}{lXrrrr}
      \toprule
      \multicolumn{2}{c}{\multirow{2.5}{*}{\textbf{Origin}}} & \multicolumn{2}{c}{\textbf{Train}} & \multicolumn{2}{c}{\textbf{Dev}} \\ \cmidrule(lr){3-4}\cmidrule(lr){5-6}
      & & \multicolumn{1}{c}{\textbf{\# Spk}} & \multicolumn{1}{c}{\textbf{\# Utt}} & \multicolumn{1}{c}{\textbf{\# Spk}} & \multicolumn{1}{c}{\textbf{\# Utt}}  \\ \midrule[.5pt]
      \textbf{Real} & AISHELL-3~\cite{shi2020aishell3} & 83 & 26,554 & 38 & 8,914 \\
      \textbf{Fake} & HAD~\cite{Yi2021Half} & 83 & 26,554 & 33 & 8,914 \\
     \bottomrule
    \end{tabularx}
  \end{subtable}

  \vspace{\baselineskip}

  \begin{subtable}{\linewidth}
    \centering
    \caption{Testing sets.}\label{tab:datasets:track-2:test}
    \begin{tabularx}{\linewidth}{lXrr}
      \toprule
      \multicolumn{2}{c}{\textbf{Origin}} & \multicolumn{1}{c}{\textbf{\# Spk}} & \multicolumn{1}{c}{\textbf{\# Utt}}  \\ \midrule[.5pt]
      \multirow{7.5}{*}{\textbf{Real}} & AISHELL-3 & 50 & 767 \\
      & AISHELL-1 & 200 & 2,482 \\
      & THCHS-30 & 30 & 439 \\
      & AliMeeting & 500 & 4,953 \\
      & Phone recording & 150 & 2,419 \\
      & Mobvoi & 100 & 8,940 \\ \addlinespace
      & \textbf{Total} & 1,030 & 20,000 \\ \midrule
      \multirow{4.5}{*}{\textbf{Fake}} & Once & 18 & 10,000 \\
      & Twice & 26 & 10,000 \\
      & Speech edit &  148 & 10,000 \\ \addlinespace
      & \textbf{Total} &  148 & 30,000\\ \bottomrule
    \end{tabularx}
  \end{subtable}
\end{table}

\begin{table}
  \centering
  \caption{Dataset for the AR track (Track 3)}\label{tab:datasets:track-3}
  \begin{subtable}{\linewidth}
    \centering
    \caption{Training and development sets.}\label{tab:datasets:track-3:train}
    \begin{tabularx}{\linewidth}{llXrrrr}
      \toprule
      \multicolumn{3}{c}{\multirow{2.5}{*}{\textbf{Origin}}} & \multicolumn{2}{c}{\textbf{Train}} & \multicolumn{2}{c}{\textbf{Dev}} \\ \cmidrule(lr){4-5}\cmidrule(lr){6-7}
      & & (Label) & \multicolumn{1}{c}{\textbf{\# Spk}} & \multicolumn{1}{c}{\textbf{\# Utt}} & \multicolumn{1}{c}{\textbf{\# Spk}} & \multicolumn{1}{c}{\textbf{\# Utt}}  \\ \midrule[.5pt]
      \textbf{Real} & AISHELL-3 & \texttt{6} &  27 & 3,200 & 10 & 1,200 \\\midrule
      \multirow{6.5}{*}{\textbf{Fake}} & Aliyun & \texttt{0} & 10 & 3,200 & 2 & 1,200 \\
      & DataBaker& \texttt{1}  & 10 & 3,200 & 2 & 1,200 \\
      & Aispeech& \texttt{2}  & 10 & 3,200 & 2 & 1,200 \\
      & HiFiGAN & \texttt{3} & 40 & 3,200 & 10 & 1,200 \\
      & WaveNet & \texttt{4} & 40 & 3,200 & 10 & 1,200 \\
      & World & \texttt{5} & 40 & 3,200 & 10 & 1,200 \\ \addlinespace
      & \textbf{Total} & & 150 & 19,200 & 36 & 7,200 \\
     \bottomrule
    \end{tabularx}
  \end{subtable}

  \vspace{\baselineskip}

  \begin{subtable}{\linewidth}
    \centering
    \caption{Testing sets.}\label{tab:datasets:track-3:test}
    \begin{tabularx}{\linewidth}{llXlrr}
      \toprule
      \multicolumn{3}{c}{\textbf{Origin}} & \textbf{Condition} & \multicolumn{1}{c}{\textbf{\# Spk}} & \multicolumn{1}{c}{\textbf{\# Utt}}  \\ \midrule[.5pt]
      \multicolumn{2}{c}{\multirow{3.5}{*}{\textbf{Real}}} & \multirow{3.5}{*}{AISHELL-3} & Clean & 34 & 4,008 \\
      & & & Noisy & 44 & 3,500 \\
      & & & Compressed & 20 & 2,999 \\\addlinespace
      & & \textbf{Total} & & 98 & 10,507 \\ \midrule
      \multirow{24}{*}{\textbf{Fake}} & \multirow{21}{*}{\textbf{Known}} & \multirow{3}{*}{Aliyun} & Clean & 4 & 4,008 \\
      & & & Noisy & 13 & 3,500 \\
      & & & Compressed & 10 & 2,004 \\\cmidrule(lr){3-6}
      & & \multirow{3}{*}{DataBaker} & Clean & 4 & 4,008 \\
      & & & Noisy & 4 & 3,500 \\
      & & & Compressed & 10 & 2,966 \\\cmidrule(lr){3-6}
      & & \multirow{3}{*}{Aispeech} & Clean & 4 & 4,008 \\
      & & & Noisy & 6 & 1,822 \\
      & & & Compressed & 10 & 1,339 \\\cmidrule(lr){3-6}
      & & \multirow{3}{*}{HiFiGAN} & Clean & 34 & 4,008 \\
      & & & Noisy & 43 & 3,500 \\
      & & & Compressed & 41 & 2,953 \\\cmidrule(lr){3-6}
      & & \multirow{3}{*}{WaveNet} & Clean & 34 & 4,008 \\
      & & & Noisy & 47 & 3,500 \\
      & & & Compressed & 43 & 2,883 \\\cmidrule(lr){3-6}
      & & \multirow{3}{*}{World} & Clean & 34 & 4,008 \\
      & & & Noisy & 50 & 3,500 \\
      & & & Compressed & 30 & 2,999 \\\cmidrule{2-6}
      & \multirow{3}{*}{\textbf{Unknown}} & \multirow{3}{*}{Baidu} & Clean & 4 & 4,008 \\
      & & & Noisy & 5 & 3,500 \\
      & & & Compressed & 10 & 2,961 \\\cmidrule{2-6}
      & \multicolumn{3}{l}{\textbf{Total (known + unknown)}} & 440 & 68,983 \\
     \bottomrule
    \end{tabularx}
  \end{subtable}
\end{table}

\subsection{Track 2: Manipulation region location (RL)}\label{ssec:challenge-outline:rl}


The RL track is designed to evaluate the ability of participants to detect and locate the manipulation regions within partially fake audio, therefore the dataset used in this track (see Table~\ref{tab:datasets:track-2}) contains both completely real audio samples as well as samples spliced with fake audio or real audio from a different recording of the same speaker. The spliced regions either contain a named entity or a word that is semantically an antonym of the original word. For simplicity, each edited sentence contains at most two spliced regions.

As such, the setup of the RL track dataset aims to prompt the generalization of models to detect manipulation regions in real-world scenarios, where the data may be more varied and complex than in controlled settings. By providing a diverse and challenging dataset, we aim to push the development of models that can perform well not only on the data they are trained on but also on the unpredictable and diverse data they will encounter in practical applications.

\subsubsection{Training and development sets}

The training and development sets (see Table~\ref{tab:datasets:track-2}(a)) contain genuine audio sourced from the training and development sets of the AISHELL-3 dataset and partially fake audio sourced from the training and development sets of the HAD dataset~\cite{Yi2021Half}. The choice of these datasets is motivated by the need to balance the complexity of the task with the availability of high-quality data. As the HAD dataset contains partially fake audio obtained by splicing genuine audio from AISHELL-3 with fake audio~\cite{Yi2021Half}, it provides a more-or-less controlled environment for training and validating models to detect manipulation regions.

\subsubsection{Testing set}

The testing set is built to contain a more diverse and realistic set of scenarios than the training and development sets, reflecting the broader range of conditions that models are likely to face in real-world applications. By including a wider variety of sources and manipulation techniques, some of which may have been real audio spliced with other real audio, the testing set provides a more challenging benchmark for evaluating the performance of manipulation region location models.

The testing set of Track 2 (see Table~\ref{tab:datasets:track-2}(b)) contains 1,030 speakers and 20,000 utterances for genuine audio from various sources like AISHELL-1, AISHELL-3, THCHS-30, Mobvoi, AliMeeting and Phone recordings. Despite the overlap in the sources of genuine audio between Track 1.2 and Track 2 testing sets, there is no overlap in actual utterances between the two tracks.
The fake audio samples in the testing set are AISHELL-1 and AISHELL-2 samples spliced with either same-speaker recordings or fake audio generated with the Tacotron~2--LPCNet pipeline, with one or two spliced regions per sentence, as well as utterances processed by the Speech Edit model based on CampNet~\cite{wang2022campnet}. The fake audio samples were generated using the same speakers as the genuine audio samples. Genuine utterances are then spliced by replacing certain segments with fake audio or other recordings of the same speaker, following a similar procedure as in~\cite{Yi2021Half}. These partially-fake samples total 30,000 utterances from 148 speakers.

\subsection{Track 3: Deepfake algorithm recognition (AR)}\label{ssec:challenge-outline:ar}


The objective of the AR track is to recognize the deepfake algorithm used to generate a given piece of audio. As such, the dataset for this track is designed to evaluate the ability of participants to recognize the deepfake algorithm. By including a diverse set of sources and manipulation techniques, both known and unknown, the dataset provides a benchmark for evaluating the performance of deepfake algorithm recognition models. The setup of the AR track dataset aims to prompt the generalization by the models to recognize deepfake algorithms in real-world scenarios, where the data may be more varied and complex than in controlled settings.

The dataset (see Table~\ref{tab:datasets:track-3}), is based on the datasets presented in \cite{ma2023cfad,yan2023fingerprint}
is composed of genuine utterances from the AISHELL-3 dataset and fake utterances generated by various deepfake algorithms from the AISHELL-1 and AISHELL-3 text corpora. The deepfake algorithms used to generate the fake utterances include vocoders like HiFiGAN~\cite{kong2020hifi}, WaveNet~\cite{oord2016wavenet} and World~\cite{morise2016world}, as well as commercial solutions from Aliyun\footnote{\url{https://ai.aliyun.com/nls/tts}}, DataBaker\footnote{\url{https://www.data-baker.com/specs/compose/online}}, and AISpeech\footnote{\url{https://beta.duiopen.com/openSource/technology/tts}}.

In the training and development sets, the genuine audio samples are sourced from AISHELL-3, while the fake audio samples are sourced from the aforementioned deepfake algorithms and commercial TTS platforms. The testing set, on the other hand, contains audio samples from the same sources as the training and development sets, but with different speakers. It also includes audio samples from one unknown deepfake algorithm not included in the above list:
\begin{itemize}[leftmargin=*]
  \item \textbf{Baidu\footnote{\url{https://ai.baidu.com/tech/speech/tts}}:} A deep learning-based TTS(TTS) system developed by Baidu, powered by the latest deep learning technologies. (labelled as \texttt{7})
\end{itemize}

Further augmentations were made to the testing set to emulate real-world scenarios. With the aforementioned real and fake audio samples forming the ``clean'' condition, additional conditions were created to test the robustness of the deepfake algorithm recognition models. To simulate diverse acoustic environments of the real world, various types and levels of background noise were added to clean audio samples, creating the ``noisy'' condition to test deepfake algorithm recognition under suboptimal conditions. Additionally, to reflect audio compression commonly found on social media, the Jinshi Video Assistant software\footnote{\url{https://www.drmfab.cn/zhushou/}} was used to compress clean audio, forming the ``compressed'' condition. These steps assess the algorithm's adaptability to encoding changes.

\begin{table*}
  \centering
  \caption{DSR (\%) and methods of top-performing systems in Track 1.1 (FG-G) submissions.\@ [``Aug.''~=~augmentation; ``Rep.''~=~representation; ``spec''~=~spectrogram; ``Arch.''~=~architecture; ``AR''~=~auto-regressive]}\label{tab:track-1-1-submissions}
  \setlength{\tabcolsep}{1.5mm}{
  \begin{tabular}{crlllllllll}
    \toprule
    Team & DSR ($\uparrow$) & Data Aug. & Text Rep. & Audio Rep. & Output & Arch. & Vocoder & Duration & Speaker Rep. & AR \\ \midrule
    A01~\cite{add2023_a01} & 44.97 & Noise + reverb &  Phoneme seq. & Mel spec & Mel spec & Tacotron 2 & WaveRNN &  -  & BiLSTM & AR \\
    A02~\cite{add2023_a02} & 43.63 &  -  & Phoneme seq. & Mel spec & Mel spec & FastPitch & HiFiGAN & MFA & One-hot & NAR \\
    A03~\cite{add2023_a03} & 41.48 &  -  & Context encoder & Linear spec & Waveform & Hier-TTS &  -  &  -  & One-hot & AR \\
    A05~\cite{add2023_a05} & 37.35 & Concatenation & G2P+BERT & Mel spec & Mel spec & FastSpeech 2 & HiFiGAN & Kaldi & One-hot & NAR \\
        & & of samples &          &          &          & + Conformer & & force-align & & \\
    A06~\cite{add2023_a06} & 30.69 &  - & Phoneme seq. & Linear spec & Waveform & VITS &  -  & MAS & One-hot & NAR  \\
    \bottomrule
  \end{tabular}}
\end{table*}

\begin{table*}
  \centering
  \caption{WEER (\%) and methods of top-performing systems and baselines in Track 1.2 (FG-D).\@ [``Aug.'' = augmentation]}\label{tab:track-1-2-submissions}
  \begin{tabular}{crllll}
    \toprule
    Team & WEER ($\downarrow$) & Data Aug. & Acoustic Features & Back-end Classifiers & Model Fusion \\ \midrule
    B01~\cite{add2023_b01} & 12.45 & Noise; RawBoost~\cite{Tak2022Rawboost}; copy synthesis & Wav2Vec~2.0 & AASIST (-sinc conv) &  - \\
    B02~\cite{add2023_b02} & 17.93 & Noise & Wav2Vec~2.0 & SENet; LCNN; AASIST& Weighted average\\
    B03~\cite{add2023_b03} & 22.13 & Noise & CQT spectrogram & LCNN; AASIST & Average\\
    B04~\cite{add2023_b04} & 22.45 &  -  & Wav2Vec~2.0; WavLM & VAE & Average \\
    B05~\cite{add2023_b05} & 23.17 & Noise & Wav2Vec~2.0 & LCNN &\\ \midrule
    S01~\cite{yi2023add2023} & 53.04 &  -  & LFCC & GMM &  - \\
    S02~\cite{yi2023add2023} & 66.72 &  -  & LFCC & LCNN &  - \\
    S03~\cite{yi2023add2023} & 30.35 &  -  & Wav2Vec~2.0 & LCNN &  - \\
    \bottomrule
  \end{tabular}
\end{table*}

\section{Technical Analysis}\label{sec:challenge-results}

We analyze the technical details of the top-5 performing teams of each track in this section, in order to provide insights into the strategies and techniques that led to their success and to identify common trends and best practices in the field of deepfake audio detection and analysis. In the tables are presented not only the performances of top-performing participating teams, but also the results of the baseline systems~\cite{yi2023add2023} (denoted in the form of \emph{S\#}, where \# is the baseline number). Among the baselines, \emph{S01}--\emph{S03} are the baseline systems for the FG-D sub-track, \emph{S04} is the baseline systems for the RL track, and \emph{S05}--\emph{S06} are the baseline systems for the AR track.

\subsection{Track 1.1: Generation task (FG-G)}

Although the use of VC was allowed, most teams opted for a TTS approach. The technical details of the top-5 participating teams are summarized in Table~\ref{tab:track-1-1-submissions}. Given the setup of the track, where the target speaker ID is selected from the speakers within the AISHELL-3 corpus, most teams opted to use one-hot embeddings for speaker representation, instead of style embeddings. In the analysis of top-performing systems, we noticed the following:
\begin{enumerate}[leftmargin=*]
  \item \textbf{Data augmentation:} The use of data augmentation in Track 1.1 is somewhat limited compared to other tracks, and the teams that adopted it, A01~\cite{add2023_a01} and A05~\cite{add2023_a05}, used it in the process of speaker ID and characteristic modelling rather than in the audio generation process. Of those two teams, A01 uses reverberation and additive noises to augment the data, while A05 concatenates and re-splices training data belonging to the same speaker ID. This might be due to the fact that the AISHELL-3 dataset is already quite large, and the use of data augmentation might not be as crucial as in other tasks.
  \item \textbf{Text representation:} Most teams opted for phoneme sequence as the text representation, which is a common choice for TTS systems, as it is seen to be closer to the phonetic representation of speech, and thus more suitable for speech synthesis, as well as being more robust to spelling errors and out-of-vocabulary words. A notable exception to this trend is Team A03~\cite{add2023_a03}, which used a context encoder to encode the input text into a fixed-length vector. Another deviation from this is Team A05~\cite{add2023_a05}, who used a grapheme-to-phoneme model to convert the input text into phoneme sequence, and used BERT for disambiguating homographs. This approach is more robust to spelling errors and out-of-vocabulary words, as it can handle unseen words by converting them into phonemic representations which are more directly related to the speech signal and also present in the training data in some form.
  \item \textbf{Architectures:} The use of non-autoregressive models like FastSpeech~2 is common in this track, which allows for variations in generation pipelines in the form of duration models. Aside from using models that are non-autoregressive, most teams adopt approaches with intermediate audio representations like Mel spectrogram, which are then passed through vocoder models to generate the final audio waveform. It is worth noting, however, that A06~\cite{add2023_a06} used the VITS architecture, which is fully end-to-end, with its own alignment module of Monotonic Alignment Search, that allows for the generation of audio waveform directly from phoneme sequence by using a variational autoencoder-like encoder-decoder architecture to directly decode into audio waveform. This approach is more computationally efficient and allows for faster inference times, as it does not require the intermediate step of generating Mel spectrograms.
\end{enumerate}

\begin{table*}
  \centering
  \caption{Scores (\%), sentence accuracies $A_s$ (\%), segment-wise F\textsubscript{1} scores (\%) and methods of top-performing systems and baseline in Track~2 (RL).\@ [``Aug.'' = augmentation]}\label{tab:track-2-submissions}
  \begin{tabular}{crrrllll}
    \toprule
    Team & Score ($\uparrow$) & $A_s$ ($\uparrow$) & $F_{1s}$ ($\uparrow$)& Data Aug. & Acoustic Features & Back-end Classifiers & Model Fusion \\ \midrule
    C01~\cite{add2023_c01} & 67.13 & 82.23 & 60.66 &  Noise; reverb & Wav2Vec~2.0 & ResNet--Transformer--LSTM &  - \\
    C02~\cite{add2023_c02} & 62.49 & 80.91 & 54.60 & Noise; re-splicing & Spectrogram & RCNN-BLSTM &  - \\
    C03~\cite{add2023_c03} & 62.42 & 79.56 & 54.50 & Noise; reverb& Log-mel spectrogram & RCNN &  - \\
    C04~\cite{add2023_c04} & 59.62 & 78.16 & 51.67 &  -  & Wav2Vec~2.0 & Transformer--BLSTM &  - \\
    C05~\cite{add2023_c05} & 59.12 & 74.52 & 52.53 & Noise & Raw, Wav2Vec~2.0 & AASIST, FC layer &  Weighted segment-wise \\ \midrule
    S04~\cite{yi2023add2023} & 42.25 & & &  -  & LFCC & LCNN &  - \\
    \bottomrule
  \end{tabular}
\end{table*}

\begin{table*}
  \centering
  \caption{$F_1$ (\%) and methods of top-performing systems and baselines in Track~3 (AR).\@ [``Aug.'' = augmentation]}\label{tab:track-3-submissions}
  \setlength{\tabcolsep}{1.3mm}{
  \begin{tabular}{crlllll}
    \toprule
    Team & $F_1$ ($\uparrow$) & Data Aug. & Acoustic Features & Back-end Classifier & Model Fusion & OSR Method \\ \midrule
    D01~\cite{add2023_d01} & 89.63 & Noise, reverb, CutMix & STFT, Wav2Vec~2.0 & SENet, LCNN-LSTM, TDNN & Weighted average & kNN \\
    D02~\cite{add2023_d02} & 83.12 & Noise, reverb & log-mel filterbank; & ResNet34SimAM-ASP, ResNet34-GSP, & Average (score) & maximum\\ &&&log-spec& ResNet34SE-ASP, ECAPA-TDNN-ASP, & & similarity\\ &&&& LCNN,  AASIST-SAP, wav2vec-ECAPA, \\ &&&& wavlm-ECAPA  \\
    D03~\cite{add2023_d03} & 75.41 & Noise, reverb, mixup & Wav2Vec~2.0 & ECAPA-TDNN &  -  & threshold \\
    D04~\cite{add2023_d04} & 73.55 &Noise, random sampling, & log mel spec, WavLM & ResNet101-Temporal-Frequency & Weighted average & threshold \\ &&  time stretching, time & & Transformer (TFT)\\ && masking, freq. masking \\
    D05~\cite{add2023_d05} & 73.52 & Noise, remove silence & raw, LFCC, HuBERT & RawNet2, SE-Res2Net50, HuBERT & Label fusion; & manifold-based \\ &&&&& average (score); &  multi-model \\ &&&&& concat (feature)& fusion \\ \midrule
    S05~\cite{yi2023add2023} & 53.50 &  -  & LFCC & ResNet &  -  & threshold\\
    S06~\cite{yi2023add2023} & 54.16 &  -  & LFCC & ResNet &  -  & OpenMax\\
    \bottomrule
    \end{tabular}
  }
\end{table*}

\subsection{Track 1.2: Detection task (FG-D)}

The technical details of the top-5 participating teams are summarized in Table~\ref{tab:track-1-2-submissions}. Given the binary nature of the task, it is interesting to note that Team B04 opted to use a variational autoencoder (VAE) as their back-end classifier.
In the analysis of top-performing systems, we noticed the following:
\begin{enumerate}[leftmargin=*]
  \item \textbf{Data augmentation:} The use of data augmentation on the training data in Track 1.2 is widespread and has found much success, with most teams opting to use MUSAN~\cite{snyder2015musan} noise augmentation. The use of reverberation is also popular, with most teams opting to add reverberation to their training data to improve the robustness of their models. This is likely due to the fact that the testing set contains a wide variety of both real and fake audio samples with different conditions, and thus data augmentation may be crucial to ensure that the model is robust to these different conditions. Interestingly, Team B01~\cite{add2023_b01} opted to furthermore use copy synthesis as one of their data augmentation methods to obtain more training data from the AISHELL-3 corpus, and given the success of their system, this approach may have been effective in the task.
  \item \textbf{Acoustic Featuress:} Most teams opted to incorporate acoustic feature extractions into their systems, and Teams B01~\cite{add2023_b01}, B02~\cite{add2023_b02} and B03~\cite{add2023_b03} notably tweaked the AASIST backend, originally an end-to-end model, to accomodate the use of acoustic features like wav2vec and CQT spectrograms.
  The reliability and popularity of Wav2Vec~2.0~\cite{Baevski2020wav2vec2} is evident in the fact that most teams opted to use it as their acoustic feature extractor, with the notable exception of Team B03~\cite{add2023_b03}, who used CQT spectrogram. This popularity is likely due to the fact that deep embedding features are more robust to noise and reverberation, and thus may be more suitable for real-world applications. The promising performance of deep embedding features in anti-deepfake systems is also evidenced in the relatively high performance of the baseline system \textbf{S03} in the FG-D track, which employs Wav2Vec~2.0 as its acoustic feature extractor, and is the best-performing baseline system in this track.
  \item \textbf{Back-end classifier:} LCNN-based~\cite{Wu2020LCNN} models are popular among participating teams due to its efficiency and performance. Notably, AASIST~\cite{jung2022aasist}, in part based on RawNet2~\cite{Jung2020RawNet2} and incorporating graph attention networks, is also featured in many submissions by participants of Track 1.2, with multiple top-performing teams opting to use AASIST as one of their back-end classifiers, if not the only one. We thus feel that AASIST is representative of progress in anti-deepfake system development. Interestingly, however, participants often favour the use of wav2vec deep embeddings in conjunction with the AASIST backend by removing the sinc-conv layers, presumably to benefit from both the expressivity of wav2vec features as well as the performance of the AASIST model.
  \item \textbf{Model fusion:} While a significant number of teams opted for a single-model approach, many teams opted for a multi-model fusion approach. Among those latter teams, the most popular approach is simple averaging, with the notable exception of Team B02~\cite{add2023_b02}, who used a weighted average of the predictions from SENet~\cite{hu2018squeeze}, LCNN~\cite{Wu2020LCNN} and AASIST~\cite{jung2022aasist}. This approach is likely effective due to the fact that different models may have different strengths and weaknesses, and thus a multi-model fusion approach may be able to combine the strengths of different models to achieve better performance while mitigating their respective weaknesses.
\end{enumerate}

\subsection{Track 2: Manipulation region location (RL)}

The technical details of the top-5 participating teams are summarized in Table~\ref{tab:track-2-submissions}. Like in the FG-D task, Wav2Vec~2.0-extracted features are popular among the top-performing teams. As for back-end classification models, LSTM-based models are popular, with the notable exception of Team C05, who used an AASIST-based model. The use of data augmentation is also widespread, with most teams opting to use MUSAN for noise augmentation.
In the analysis of top-performing systems, we noticed the following:
\begin{enumerate}[leftmargin=*]
  \item \textbf{Data augmentation:} As with other tracks of the challenge, data augmentation sees widespread use in the RL track, with most teams using additive noise augmentation. Interestingly, Team C02~\cite{add2023_c02} furthermore opted to use re-splicing as one of their data augmentation methods, which is well suited for this task, given that the RL track is concerned with the detection of manipulation regions, and resplicing allows for the creation of more training data with different manipulation regions.
  \item \textbf{Acoustic features:} Like in the FG-D task, most teams opted to use Wav2Vec~2.0 for acoustic feature extraction, with the notable exception of Teams C02~\cite{add2023_c02} and C03~\cite{add2023_c03}, who used spectrograms as acoustic feature representation. This popularity of Wav2Vec~2.0 further evidences the reliability of deep embedding features in anti-deepfake systems. In addition, it is worth noting that Team C05~\cite{add2023_c05} uses both raw waveform as well as wav2vec deep embedding features, an approach that may be more effective in allowing the backend model to both capture the information contained in the raw waveform as well as the deep embedding features, and thus may be more effective in capturing the diverse acoustic conditions of deepfake audio in the wild.
  \item \textbf{Back-end classifiers:} Given the sequential nature of the RL task, where a real/fake prediction must be given for each segment, and notably due to the fact that audio waveforms are temporal in nature, the use of LSTM-based recurrent models to capture the temporal sequential informations is a popular approach among the top-performing teams, with the notable exception of Team C05~\cite{add2023_c05}, who used an AASIST-based model. Judging by the ranking, however, it is evident that LSTM-based recurrent models may still be more effective and better suited for the recurrent nature of the RL task, as Team C05, who used a fusion of AASIST-based model and FC layer classifier, is the only team in the top-5 that did not use an LSTM-based model, and the performance is slightly lower than that of the other teams.
  \item \textbf{Model fusion:} The use of model fusion is not as widespread in Track 2 compared to other tracks, with only C05 opting to use a weighted segment-wise average of the predictions from their models.
\end{enumerate}

\subsection{Track 3: Deepfake algorithm recognition (AR)}

The technical details of the top-5 participating teams are summarized in Table~\ref{tab:track-3-submissions}. Given the particular nature of open-set recognition (OSR), the out-of-distribution detection method is separately listed in the last column. In the analysis of top-performing systems, we noticed the following:
\begin{enumerate}[leftmargin=*]
  \item \textbf{Data augmentation:} The augmentation of training data to introduce more acoustic conditions, especially with noise and reverberation, is widespread among the top-performing teams. Most teams opted to use MUSAN for noise augmentation, and a significant number of teams also opted to add reverberation to their training data to improve the robustness of their models. In addition, the top-performing teams also opted to use other methods to further augment their training data, including CutMix~\cite{yun2019cutmix} and random sampling, time stretching, and time masking. This is likely due to the fact that the AR task is concerned with the recognition of deepfake algorithms, and thus the use of different data augmentation methods may be more effective in capturing the diverse acoustic conditions of deepfake audio in the wild.
  \item \textbf{Acoustic features:} Compared to Tracks 1.2 and 2, participants of Track 3 are more diverse in their choice of acoustic features. While a number of teams used Wav2Vec~2.0, a significant number of teams also opted to use log-Mel spectrogram, STFT spectrogram, and even HuBERT-based features. This is likely due to the fact that the AR task is concerned with the recognition of deepfake algorithms, beyond the binary real/fake classification, and thus a more diverse set of acoustic features may be more suitable, without a clear-cut winner. This also signals the prospect of additional research into the development of acoustic features for anti-deepfake systems.
  \item \textbf{Back-end classifiers:} ResNet-based~\cite{He2015Resnet} models (including Res2Net~\cite{gao2019res2net}) are popular among the top-performing teams, with the notable exception of Team D03~\cite{add2023_d03}, who used a single ECAPA-TDNN-based model~\cite{desplanques2020ecapa}. The use of ECAPA-TDNN is likely due to the fact that ECAPA-TDNN is a popular model for speaker recognition, and the philosophy of speaker recognition may be applicable to the AR task, which is also a recognition task, albeit of deepfake algorithms instead of speakers. The popularity of ResNet-based models is likely due to the fact that they are relatively simple and efficient, and thus may be more suitable for real-world applications.
  \item \textbf{Model fusion:} The averaging of scores is still popular, but the use of label fusion and feature fusion (in the form of concatenation) also see its use. This is likely due to the fact that the AR task is concerned with the recognition of deepfake algorithms, and thus the use of different models, each of which may be better suited for a particular aspect of the task, may be more effective than a single-model approach. The use of manifold-based multi-model fusion by Team D05~\cite{add2023_d05} is particularly interesting, as it allows for the pooling of information from multiple models to make a more informed decision, and thus may be more effective in capturing the diverse acoustic conditions of deepfake audio in the wild.
  \item \textbf{OSR methods:} Given the nature of the AR task and its practicality in real-world applications, as well as the presence of an unknown class, participating teams are required to use an out-of-distribution detection method to detect unknown classes. Thresholding remains the most popular method, to decent success. Other methods, including OpenMax~\cite{bendale2016towards} and manifold-based multi-model fusion, are also used. These methods are effective in detecting unknown classes, likely because they allow for the evaluation of prediction confidence and the pooling of information from multiple models, respectively, in order to make a more informed decision.
\end{enumerate}

\section{Future Directions}\label{sec:limitations}

The ADD 2023 challenge has provided a platform for researchers to develop new technologies to combat deepfake audio; However, there are several limitations that should be addressed in future research efforts. We identify these limitations and suggest future directions for research in the field of deepfake audio detection and analysis.

\begin{itemize}[leftmargin=*]
\item \textbf{Coping with unseen deepfake technologies and adversarial attacks:} The rapid development of audio deepfake generation and adversarial attack technologies, like VALL-E, GPT-4o, VISinger and DiffSinger, etc., brings critical challenges to current existing detection methods. In response to these challenges, deepfake audio generation and detection tasks are viewed as a rivalry game for participants in the ADD competitions. Despite partly improving the anti-attack ability of the detection model via fake game, there isn't sufficient adversarial dynamic beyond generation and evaluation of deepfake speech examples. Future research should develop frameworks that enable dynamic, real-time rivalry game scenarios, allowing for a more thorough exploration of defense mechanisms' effectiveness and methods like continual reinforcement learning.

\item \textbf{Improving the interpretability of discrimination:} Beyond detecting and locating manipulated regions, future research should aim to identify specific manipulation techniques used in the manipulated audio, providing a more comprehensive understanding of the manipulation process and the reasons of discrimination. Additionally, developing visualization technologies highlighting manipulation regions and enhancing manipulation traces in audio signals can help users understand the detection process and build trust in anti-deepfake technologies. 

\item \textbf{Improving generalization ability and robustness:} Although previous studies have made some attempts on audio deepfake detection and attribution, the generalization and robustness of the models are still poor. The performance of the top-performing models in the ADD competitions are very high but it will degrade significantly when evaluated on the mismatching dataset containing multiple unseen deepfake methods or unseen acoustic conditions etc. Future studies concluding unsupervised domain adaptation, open set continual learning and transfer learning can help models better generalize, making them more reliable in real-world applications. 

\item \textbf{Considering real-time processing for detection systems:} Real-time processing is critical for deploying anti-deepfake technologies in applications such as live streaming and voice-based authentication. Future work should optimize models for low-latency performance and efficient use of computational resources. Techniques like model pruning, quantization, and edge computing can be investigated to achieve these goals, ensuring that detection systems are both responsive and resource efficient. 

\item \textbf{Considering multilingual scenarios:} The majority of previously released datasets and detection models are mainly focused on single language, most of which in English and Chinese and few of them in other language like Japanese. But the applicability of anti-deepfake technologies in multilingual scenarios is essential in realistic applications. Future research should focus on developing models that can detect deepfake speech in multiple languages, ensuring that detection technologies are effective across diverse linguistic contexts.

\item \textbf{Exploring better evaluation metrics:} EER, accuracy, precision, recall and F1-score are employed as the evaluation metric in previous work. However, evaluation metrics should be designed to reflect real-world scenarios. Future research should focus on developing standardized benchmarks and evaluation protocols that simulate real-world conditions to ensure that detection technologies are both theoretically sound and practically viable. Human detection capabilities, as well as the differences between humans and machines also need to be considered for detecting and attributing deepfake audio.
\end{itemize}

\section{Conclusions}\label{sec:conclusion}

The ADD 2023 challenge aimed to spur innovation and research in detecting and analyzing deepfake speech, attracting 145 teams from 15 countries. This paper presents the challenge's dataset and provides a technical analysis of top-performing systems. Our analysis not only identifies key strengths in the approaches used but also reveals certain limitations that need to be addressed to enhance the robustness of deepfake detection technologies, notably in the areas of better coping with unseen attacks, improved interpretability and generalization abilities, real-time processing, multilingual scenarios, and better evaluation metrics. We hope that the release of the dataset and the analysis presented in this paper will inspire further research in the field of deepfake speech detection and contribute to the development of more robust and reliable anti-deepfake technologies.

\bibliographystyle{IEEEtran}
\bibliography{Mybib,refs,add2023-submit}

\begin{IEEEbiography}[{\includegraphics[width=1in,height=1.25in,clip,keepaspectratio]{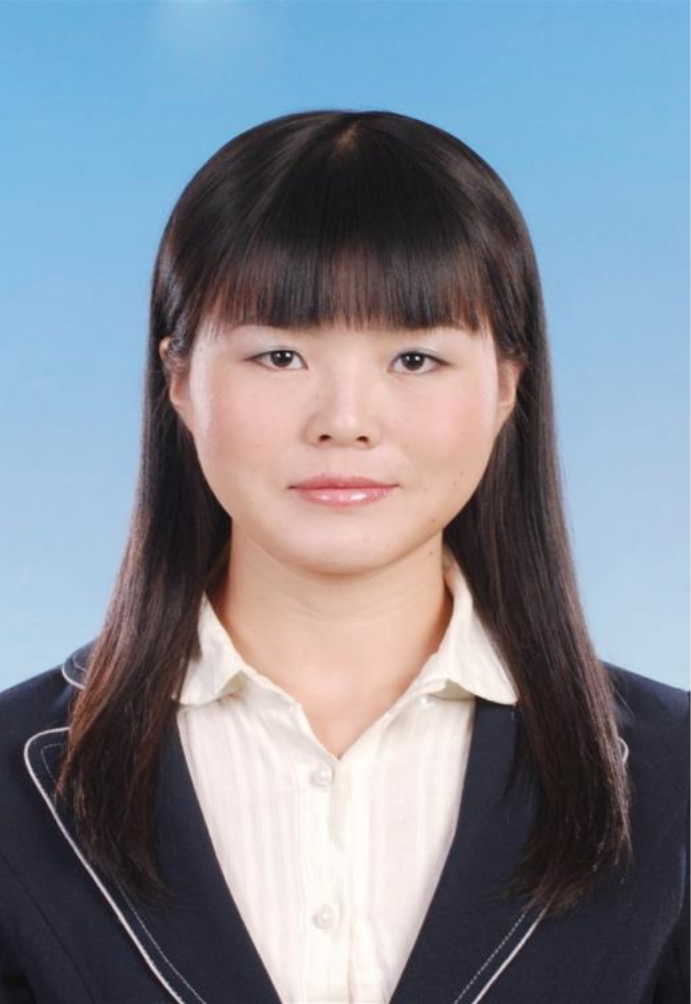}}]{Jiangyan Yi}
(Member, IEEE) received the Ph.D. degree from the University of Chinese Academy of Sciences, Beijing, China, in 2018, and the M.A. degree from the Graduate School of Chinese Academy of Social Sciences, Beijing, in 2010. During 2011 to 2014, she was a Senior R\&D Engineer with Alibaba Group. She is currently an Associate Professor with the State Key Laboratory of Multimodal Artiﬁcial Intelligence Systems, Institute of Automation, Chinese Academy of Sciences. Her research interests include speech signal processing, speech recognition and synthesis, fake audio detection, audio forensics, and transfer learning.
\end{IEEEbiography}

\begin{IEEEbiography}[{\includegraphics[width=1in,height=1.25in,clip,keepaspectratio]{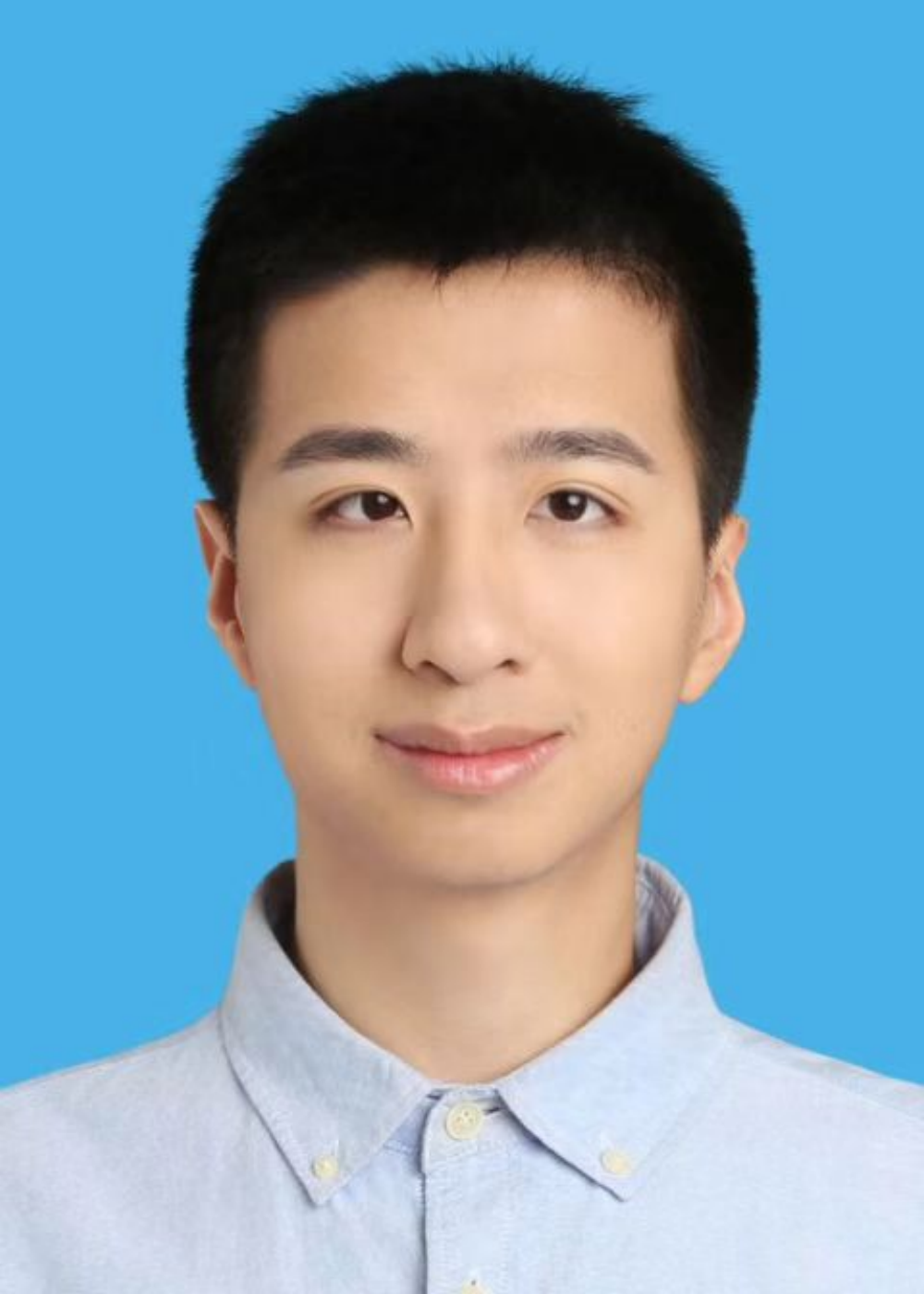}}]{Chu Yuan Zhang}
  received his M.S. degree from the Institute of Automation, Chinese Academy of Sciences, Beijing, China, in 2024, and his B.A. degree from the University of California, Los Angeles, in 2021. He is currently working toward a Ph.D. degree at the Department of Automation, Tsinghua University, Beijing, China. His research interests include speech information processing, deepfake audio detection and source attribution, and audio forensics.
  \end{IEEEbiography}

\begin{IEEEbiography}[{\includegraphics[width=1in,height=1.25in,clip,keepaspectratio]{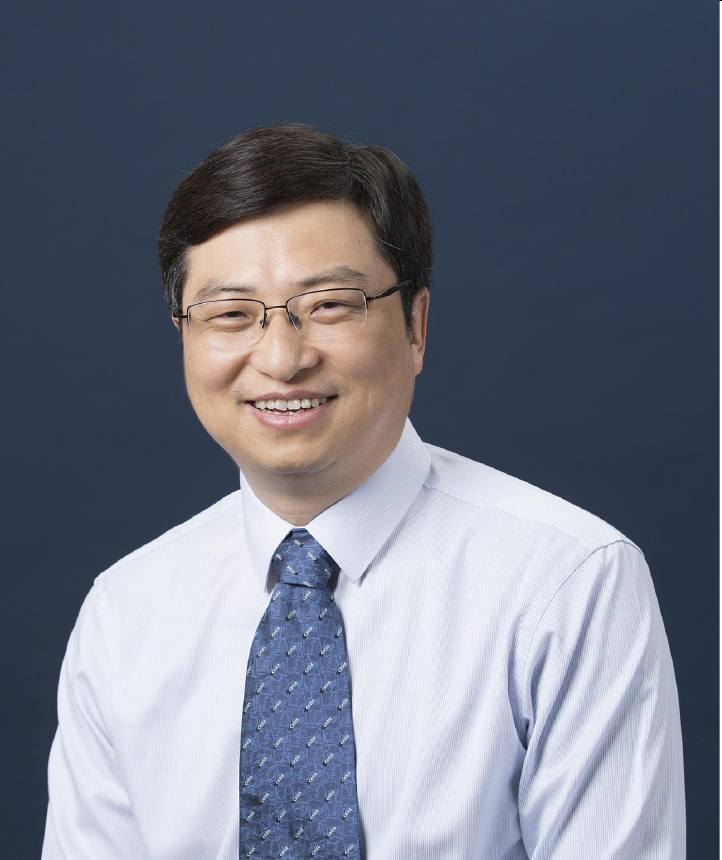}}]{Jianhua Tao}
  (Senior Member, IEEE) received the Ph.D. degree from Tsinghua University, Beijing, China, in 2001, and the M.S. degree from Nanjing University, Nanjing, China, in 1996. He is currently a Professor with the Department of Automation, Tsinghua University. He has authored or coauthored more than eighty papers on major journals and proceedings including IEEE \textsc{Transactions on Audio, Speech and Language Processing}. His research interests include speech signal processing, speech recognition and synthesis, human computer interaction, multimedia information processing, and pattern recognition.
\end{IEEEbiography}

\begin{IEEEbiography}[{\includegraphics[width=1in,height=1.25in,clip,keepaspectratio]{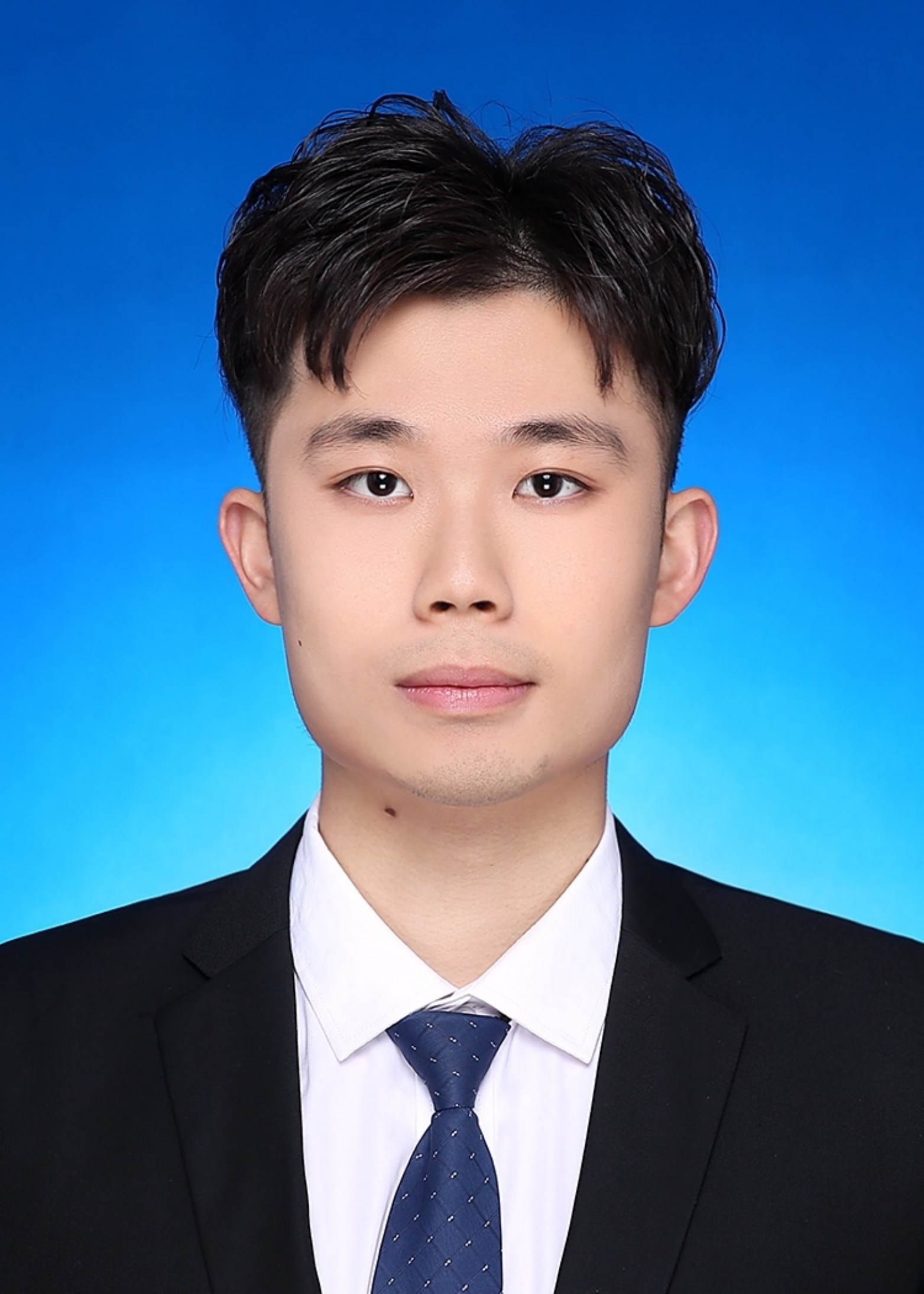}}]{Chenglong Wang}
  received the Ph.D. degree with the University of Science and Technology of China, Anhui, China in 2024. He currently works as a lecturer of the Institute of Intelligent Information Processing, Taizhou University, China. His current research interests include fake audio detection, speaker verification and identification.
\end{IEEEbiography}

\begin{IEEEbiography}[{\includegraphics[width=1in,height=1.25in,clip,keepaspectratio]{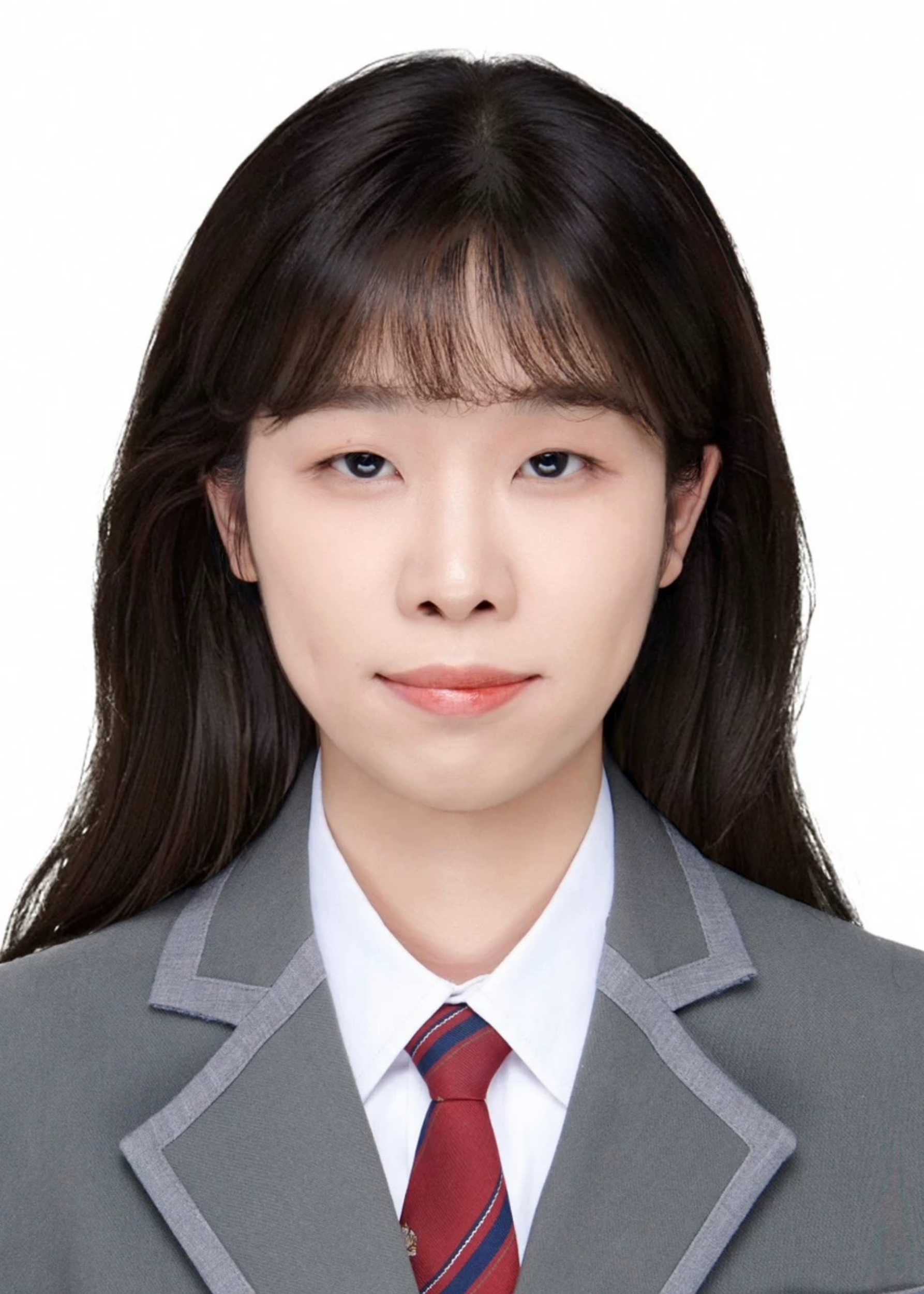}}]{Xinrui Yan}
  received her Bachelor's degree from Northeastern University in China in 2021 and obtained her Master's degree from the University of Chinese Academy of Sciences in 2024. Her current research interests include audio deepfake attribution.
\end{IEEEbiography}

\begin{IEEEbiography}[{\includegraphics[width=1in,height=1.25in,clip,keepaspectratio]{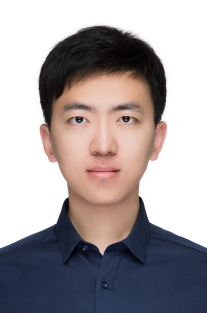}}]{Yong Ren}
  received the B.S. degree from the Department of Automation, Tsinghua University, Beijing, China, in 2020. He is currently working toward the Ph.D degree with the Institute of Automation, Chinese Academy of Sciences, Beijing, China. His current research interests include speech synthesis and audio generation.
\end{IEEEbiography}

\begin{IEEEbiography}[{\includegraphics[width=1in,height=1.25in,clip,keepaspectratio]{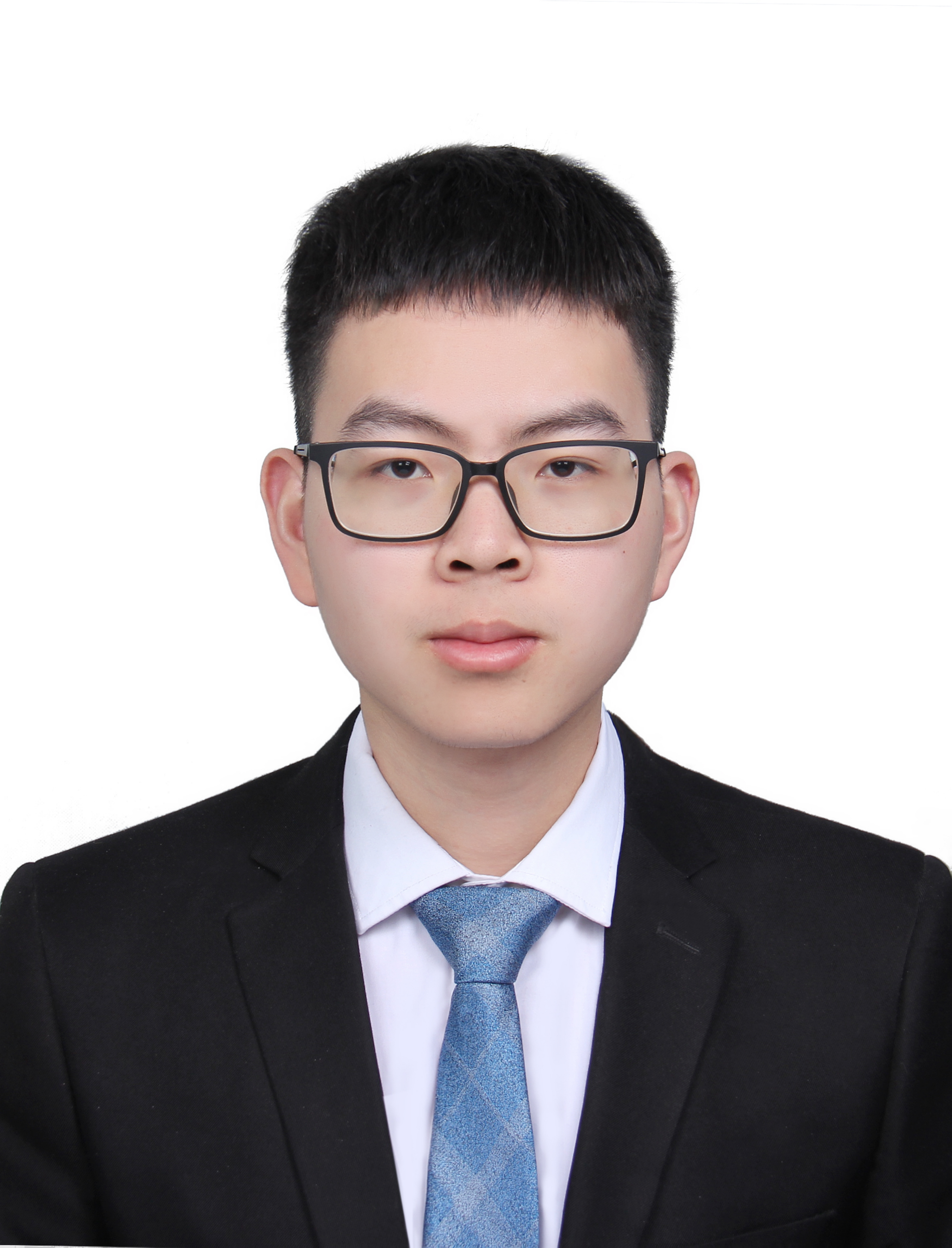}}]{Hao Gu}
  received the B.S. degree from Harbin Institute of Technology in China in 2022. He is currently pursuing his M.S. degree at Institute of Automation, Chinese Academy of Sciences in Beijing, China. His current research interest include fake audio detection.
\end{IEEEbiography}

\begin{IEEEbiography}[{\includegraphics[width=1in,height=1.25in,clip,keepaspectratio]{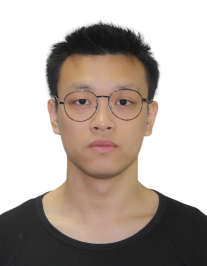}}]{Junzuo Zhou}
  reccived his B.A. deeree in Hangzhoudianzi University, Zhejiang, China, in 2023. He is currently working toward the M.S. degree with the Institute of Automation Chinese Academy of Sciences (CASIA).  His current research interests include text-to-speech.
\end{IEEEbiography}

\end{document}